\documentclass[10pt]{iopart}
\usepackage{lineno}
\usepackage{subcaption} 
\usepackage{graphicx} 
\usepackage{comment}
\usepackage{url}
\usepackage{siunitx}
\usepackage{svg}
\usepackage[colorlinks=true, linkcolor=black, citecolor=black, urlcolor=black]{hyperref}

\DeclareUnicodeCharacter{03BC}{\ensuremath{\mu}}

\begin{document}

\title[]{First Steps Toward the Development of a Straight-Line Reference Alignment System for Future Accelerators at CERN Using Pseudo-Nondiffracting Layer Beams}

\author{Martin Dušek$^{1,2}$\footnote{Corresponding author}, Sebastian Figura$^{1,3}$, Jakub Michal Polak$^1$, Solomon William Kamugasa$^1$, Dirk Mergelkuhl$^1$, Witold Niewiem$^{1, 5}$, Štěpán Kunc$^2$, Jean-Christophe Gayde$^1$, and Miroslav Šulc$^{2,4}$}
\address{$^1$The European Organization for Nuclear Research (CERN), 1211, Geneva, Switzerland}
\address{$^2$Technical University of Liberec (TUL), Studentská 1402/2, 461 17, Liberec, Czech Republic}
\address{$^3$Cracow University of Technology (CUT), ul. Warszawska 24, 31-155 Kraków, Cracow, Poland}
\address{$^4$Institute of Plasma Physics of the Czech Academy of Sciences (IPP CAS), U Slovanky 2525, 182 00, Prague, Czech Republic}
\address{$^5$Eidgenössische Technische Hochschule Zürich (ETHZ), Rämistrasse 101, 8092 Zurich, Switzerland}
\ead{\mailto{martin.dusek@cern.ch}}

\begin{abstract}
This paper presents experimental results that allow for the performance evaluation of a straight-line reference alignment system based on pseudo-nondiffracting Layer beams. Sensors, developed specifically for this system, feature four linear CMOS chips and a square aperture. This allows for simultaneous measurements along the beam path without disrupting the laser reference. Measurements, conducted over a distance of 2~m from the first to the last sensor, were compared with a laser tracker measurement to assess the sensor performance. The alignment reference generated by the Layer Beams exhibited a repeatability and reproducibility root-mean-square error (RMSE) of less than 30~\si{\micro\meter}. The relative alignment precision for a known displacement was validated with a standard deviation of 4.3~\si{\micro\meter}. The results highlight the underlying sources of noise, which are induced mainly by the cover glass, the protective film of the pixels, and the dark noise of the CMOS chips. Solutions to address these challenges are proposed. Additionally, a proof-of-concept for future development of a radiation-hard sensor utilizing optical fiber matrices is demonstrated. The RMSE of the reference position detection introduced by the fiber matrix remained below 1.3~\si{\micro\meter}. This would allow the sensor to be used reliably in high-radiation environments typical for accelerator facilities. This study serves as a foundational step toward developing a robust straight-line reference alignment system based on pseudo-nondiffracting Layer beams intended for deployment in the accelerator facilities.

\end{abstract}
\submitto{\MST (This is a preprint and it has not yet been peer reviewed.)}
\noindent{\it Keywords\/}: Accelerator Subsystems and Technologies, Alignment Technologies, Straight-line reference, Nondiffracting Beams, Caustic Beams, Layer beams

\maketitle

\section{Introduction}
\subsection{Straight-line alignment systems}
\hspace{5mm} Achieving precise geometric configurations of particle accelerator components is crucial for the optimal functioning of such complex machines. This process is known as alignment. Various measurement techniques, including polar measurements using laser trackers or leveling methods, are used to accurately determine the position of accelerator components~\cite{CERN2024}. 

So-called straight-line reference alignment systems are supplementary to the aforementioned methods and involve measuring offsets of individually aligned elements from a reference line. The key components of these systems include the alignment reference itself, reference points on the aligned objects, and sensors that measure offsets from the reference line. The requirement of modern accelerators on the straight-line reference system is to provide alignment accuracy and repeatability in the order of micrometers within a length of sliding window which is in the order of hundreds of meters long~\cite{CERN2024, niewiem2023variation}.

Wire Positioning System (WPS) and Hydrostatic Levelling System (HLS) are used in alignment monitoring in accelerator facilities around the world (Fermilab~\cite{eddy2012wire, volk2006hydro}, SLAC~\cite{wang2010prototype}). At Conseil Européen pour la Recherche Nucléaire (CERN), the straight-line alignment reference is given by the WPS's conductive wire. This gives the possibility to measure a vertical and horizontal offset from this reference using capacitive or optical sensors~\cite{MainaudDurand2012}. The vertical coordinate measurement, using capacitive or optical HLS sensors, is complementary and is used to calculate the catenary of the wire and link measurements to a geoid (local reference given by a gravitation field)~\cite{Herty2018, Biedrawa2022}. These systems can reach an accuracy of \SI{5}{\um} in the local coordinate system~\cite{MainaudDurand2012}. The WPS is susceptible to environmental perturbations such as humidity, which requires correction by the HLS. The wire is fragile and prone to breaking under its own weight. These factors limit the length of the sliding window. To overcome this, utilization of multiple overlapping reference lines can be used, however, this can limit the accuracy of the alignment procedure. 

\subsection{Optical-based straight-line reference systems}
\hspace{5mm} Optical-based straight-line reference systems are a viable alternative that could provide a straight-line reference over a longer distance than the WPS and HLS. Various optical references have been explored in other publications including a Fresnel zone plate~\cite{herrmannsfeldt1966slac, suwada2012experimental, luo2021new}, Poisson spot~\cite{feier1998poisson, griffith1990magnetic, schwalm2012straight}, Airy pattern~\cite{Iris, KEK, KEK2}, and Gaussian beam~\cite{ClicStern, batusov2015laser}. 

The optical-based methods typically suffer from the non-homogeneous refractive index of the propagation medium, which causes the beam trajectory (the alignment reference) to be non-straight. To meet the accuracy requirements, vacuum pipes are used to mitigate this effect~\cite{herrmannsfeldt1966slac, KEK}. 
An additional limitation is beam divergence. For example, in the case of the Gaussian beam, the divergence causes a loss of contrast in the transversal laser profile imaged on a camera chip. This makes the accurate measurement of the straight-line reference challenging as the dynamic range of the signal decreases. Structured laser beams (SLBs)~\cite{patent}, characterized by pseudo-nondiffracting behavior similar to one of Bessel beams, have shown promise due to their low divergence and sharply defined inner core, which allow for accurate detection of the reference even after propagating for hundreds of meters. The main advantage contrary to Bessel beams~\cite{Durnin} is their theoretically infinite propagation distance~\cite{niewiem2023variation}. 

Nevertheless, there are practical challenges when using SLB in a simultaneous multi-point straight-line reference system. The CMOS or CCD chip, which detects the position of the reference, obstructs the beam and prevents it from propagating further. This means that a system with mechanical moving shutters must be employed to move the chip in and out of the beam path. This can increase the cost and maintenance needs of the system. It also increases the time needed for the measurement. If the alignment reference is unstable during the measurement the accuracy will be affected. This can be caused by an instability in the position of the SLB generator, or due to a non-homogeneous distribution of the refractive index of the propagation medium. Both of these effects would lead to a change in the beam trajectory.

Pellicle beam splitters can be used to mitigate this problem. However, the non-parallelism of the beam splitter surfaces can cause significant angular deviation of the beam over distances of hundreds of meters~\cite{Reichold2006s}. This effect can spoil the straightness and quality of the reference line and necessitates careful calibration. 

Layer Beams (LBs) offer a promising alternative to overcome the limitation of simultaneous multi-point measurements. LBs are pseudo-nondiffracting optical beams generated using cylindrical lenses~\cite{duvsek2024generation}. Unlike rotationally symmetric SLBs, LBs exhibit reflectional symmetry and produce parallel intensity lines. This allows for simultaneous multi-point measurements without fully obstructing the beam path. This capability significantly improves measurement efficiency and accuracy by enabling simultaneous detection of multiple points along the beam path. The structured pattern of LBs allows for accurate detection of the reference even after propagating for hundreds of meters. The principle of the alignment system based on LBs will be described in greater detail in the following text.

\subsection{Contribution of the Study}
\hspace{5mm} This paper presents experimental results obtained from LBs-based alignment system using specifically developed hollow sensors. The experiments evaluate the system performance over a relatively short distance of 2~m by comparing it to laser tracker measurements. Several underlying issues were identified during initial sensor implementation and solutions for future development are presented.

Furthermore, recognizing the harsh radiation environment typical for accelerator tunnels, a proof-of-concept for radiation-hard sensors is presented utilizing optical fiber matrices for signal transmission away from radiation zones. This innovation would ensure the long-term reliability of sensor electronics by protecting them from radiation-induced damage.

The presented work serves as an essential step toward developing a robust long-distance alignment system, intended for deployment in the accelerator facilities at CERN. 

\section{Methodology}

\subsection{Layer beams}
\hspace{5mm} LBs belong to the family of so-called pseudo-nondiffracting optical beams. They are formed by the superposition of layers of two symmetrical Airy-like beams~\cite{Airy, AiryObservation} propagating with opposing transversal acceleration directions. This results in a structured transversal intensity profile consisting of parallel lines. The structured pattern of the transversal intensity distribution and low divergence facilitate the accurate detection of the alignment reference even after propagation over a distance of 140~m~\cite{dusekLBipac}.

\begin{figure}[!ht]
\centering
\includegraphics[trim=1cm 4cm 1cm 4cm,clip,width=\textwidth]{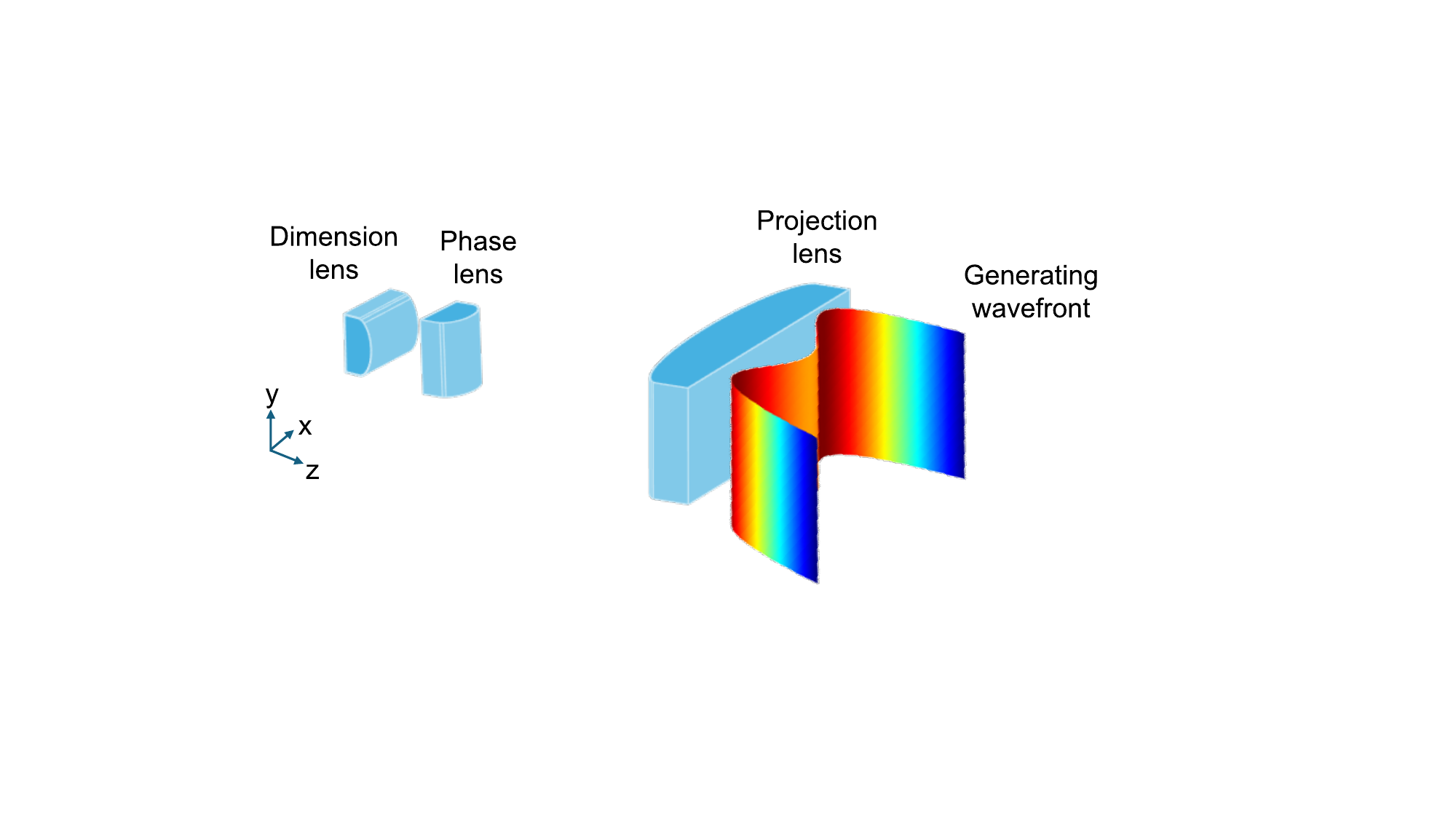}
\caption{LB generator consisting of a dimension, phase, and projection lens with an illustration of a generating wavefront that dictates the caustic properties of the beam.}
\label{LBgenerator}
\end{figure}

\begin{figure}[!ht]
\centering
\includegraphics[width=0.7\textwidth]{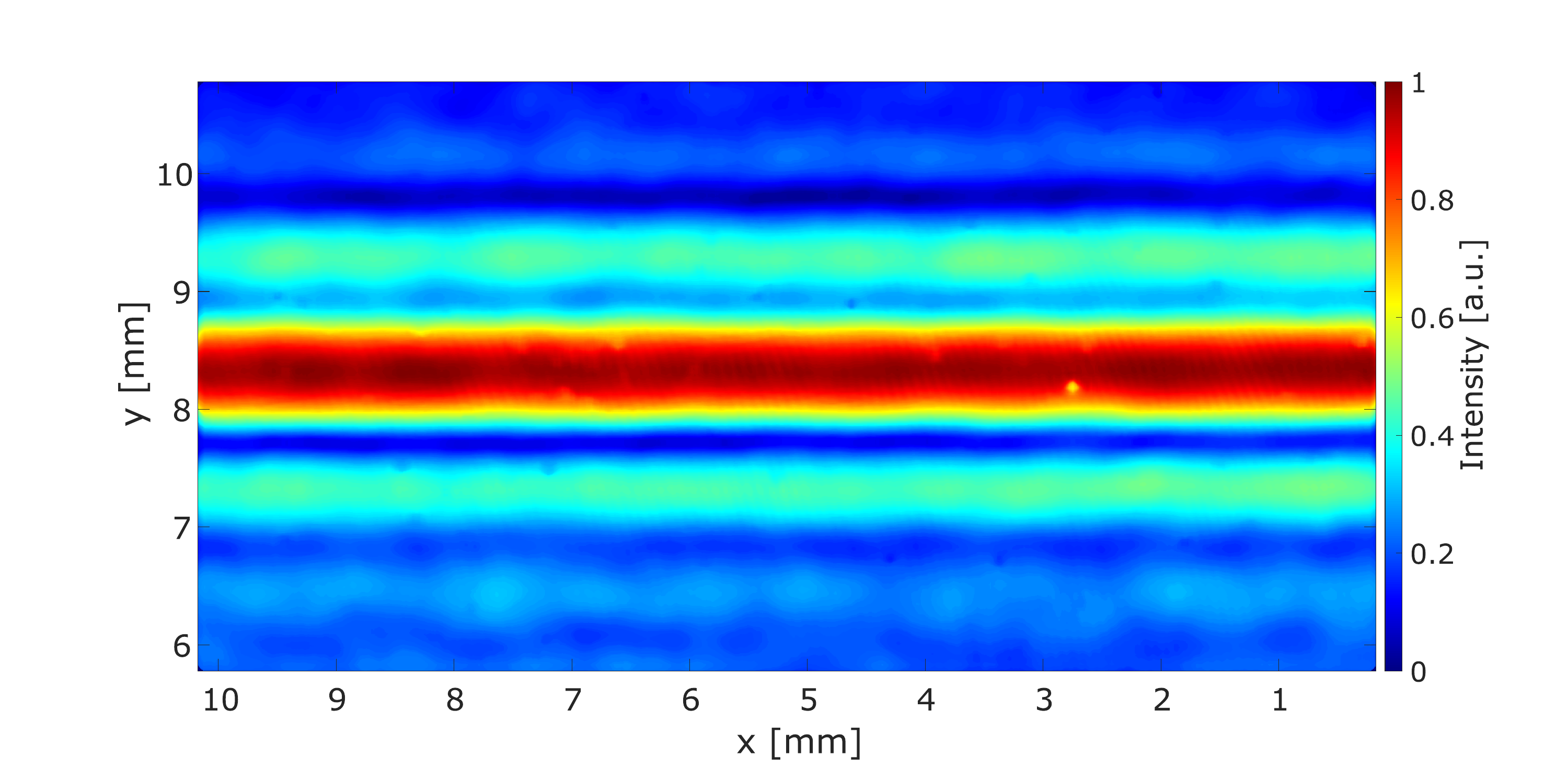}
\caption{Horizontal LB imaged on a camera at the position of the first sensor - 1.7~m away from the generator. Diffraction patterns caused by dust on the CMOS chip were removed in post-processing.}
\label{LB_profile}
\end{figure}

The generator of the LBs used in this study consists of a dimensional, phase, and projection lens. This specific combination of lenses results in the creation of a so-called generating wavefront, which in turn leads to the formation of a caustic~\cite{Stavroudis1972} that is responsible for the pseudo-nondiffracting properties of the beam as illustrated in Fig.~\ref{LBgenerator}. The dimension lens is rotated 90~degrees with respect to the other two around the optical axis, resulting in an elongation of the lines in the transversal intensity distribution caused by an increased divergence along the y-axis. The longitudinal propagation of the beam can be described by a so-called near zone and diverging zone. In the near zone the number of lines in the transversal intensity distribution changes with distance due to multiple overlapping and interfering contributions from different parts of the wavefront. In the diverging zone, the beam transversal pattern does not change with distance and only diverges. More about the generation principle and properties of the LBs can be found in our previous work~\cite{duvsek2024generation}. 

The generator of the LBs used for experiments in this study was optimized for propagation over 3.7~m. It consisted of the dimension lens with a focal length of 50~mm, and of a phase lens and projection lens, both with a focal length of 10~mm. The lenses were made of NBK-7 glass. The distance between the phase and projection lenses was optimized so that the transversal intensity distribution in the stable diverging zone of the beam consisted of a single line with small side lobes as seen in Fig.~\ref{LB_profile}. 

\subsection{Layer beam-based alignment system}
\begin{figure}[!ht]
\centering
\includegraphics[width=0.8\textwidth]{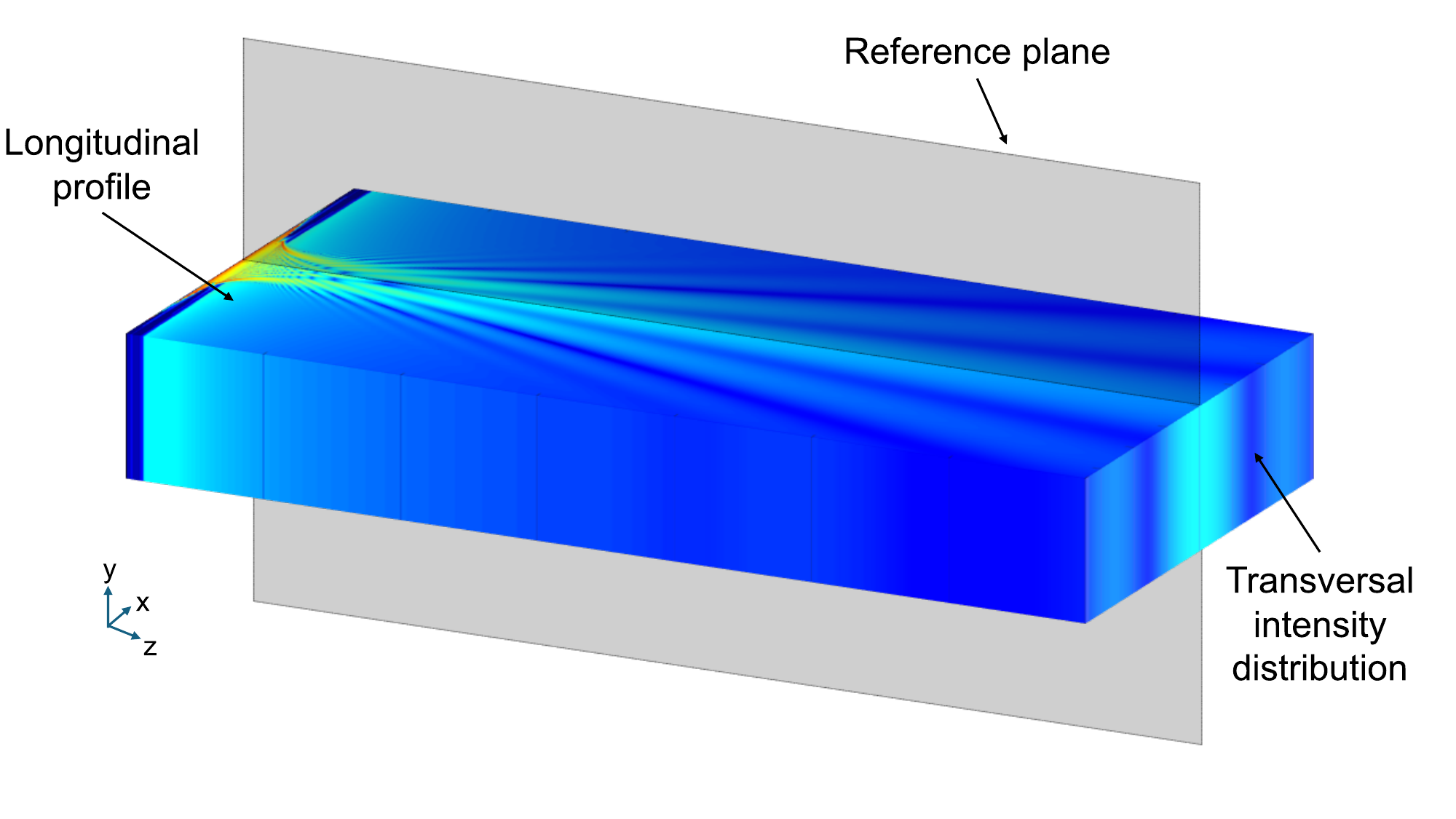}
\caption{Illustration of an alignment reference plane created by a LB.}
\label{AlignmentIllustration}
\end{figure}

\hspace{5mm} The LBs-based alignment reference relies on a reference plane rather than a reference line. This plane is generated leveraging the reflection symmetry of the transversal intensity distribution along the path of the LB, as illustrated in Fig.~\ref{AlignmentIllustration}.

In the proposed alignment system, two LBs from independent generators rotated around the optical axis with an angle close to 90~degrees were combined using a pellicle beam splitter. This approach simultaneously generates two reference planes while their intersection creates the reference line. The detection is done using a hollow sensor, which consists of four linear image sensors and an aperture. The sensor intercepts only a small portion of the beam cross-section, allowing most of it to propagate further thanks to the aperture. This ensures, together with the divergence of the beam, that the signal can be detected on multiple measurement points along the optical axis at one point in time.

\begin{figure}[!ht]
\centering
\includegraphics[width=0.8\textwidth]{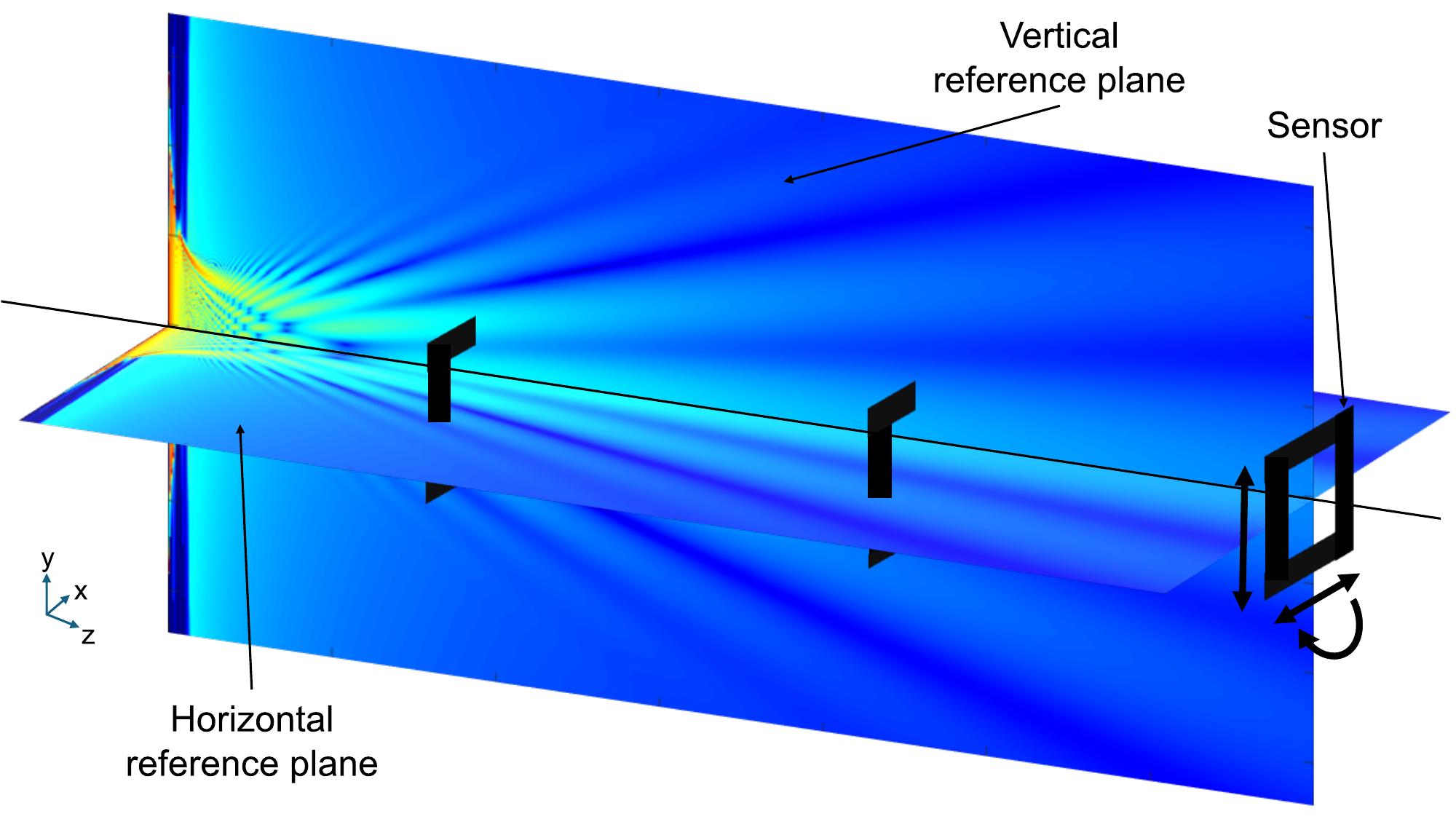}
\caption{Illustration of an alignment system based on LBs.}
\label{2dsensor_layout}
\end{figure}

Two points per reference plane are detected at each hollow sensor. A minimum of two sensors is needed to define the reference plane as by definition the plane needs to be established using a minimum of three points in space. In the alignment configuration, the first and the last sensors define the reference planes, and the positions of sensors in between them are measured with respect to this reference. Translational misalignment in both horizontal and vertical directions can be measured. Additionally, a rotational misalignment around the beam propagation axis (the roll) can also be determined as illustrated in Fig.~\ref{2dsensor_layout}. This is an advantage compared to the state-of-the-art straight-line alignment systems that can usually measure only in the translational directions.

\begin{figure}[!ht]
\centering
\includegraphics[width=0.8\textwidth]{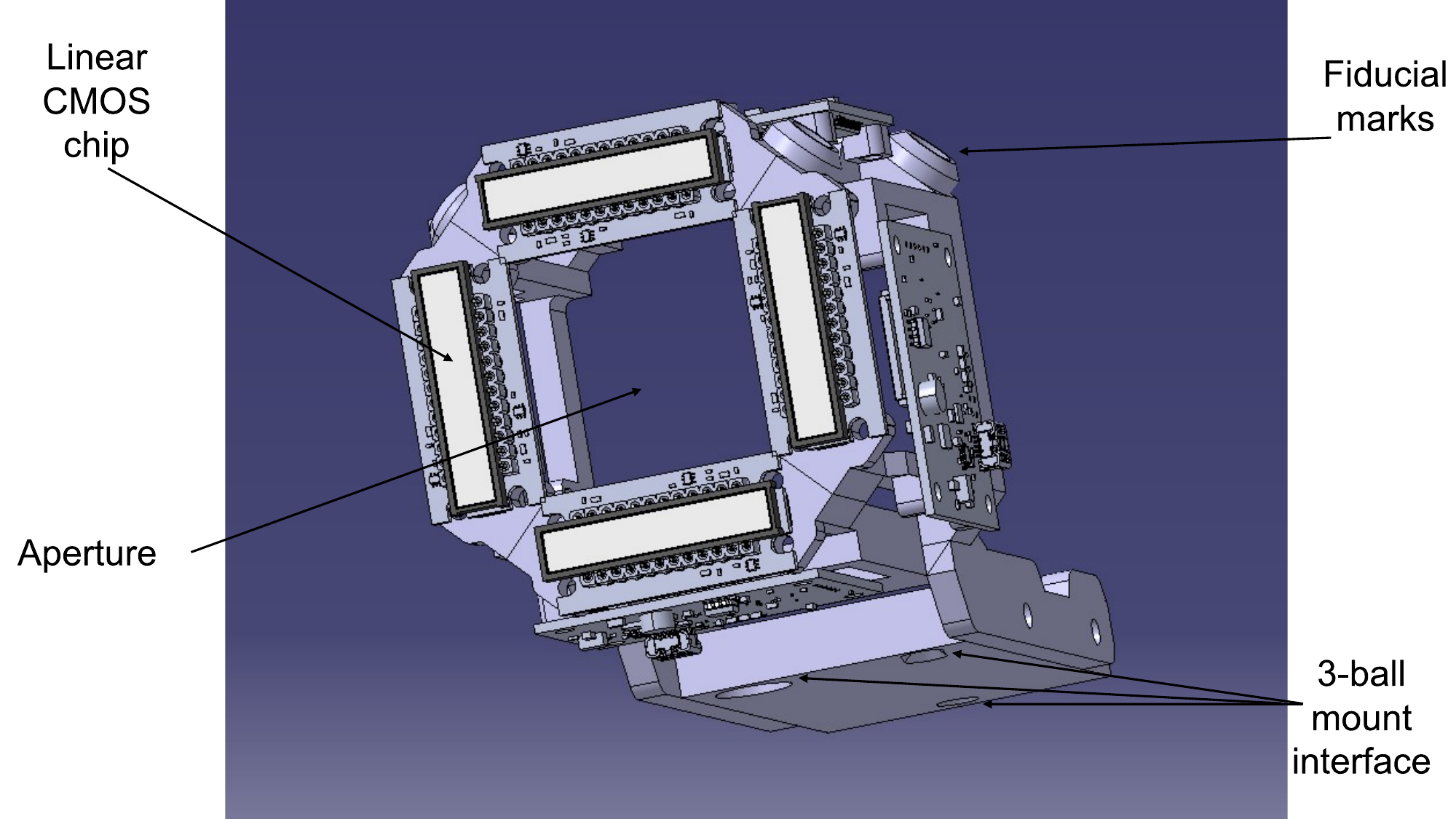}
\caption{CAD model of the hollow sensor developed for the detection of LBs. A photo of the sensor can be found in supplementary materials.}
\label{2D_sensor}
\end{figure}

Hollow sensors consisting of four linear CMOS chips placed around a squared aperture were developed for the experiment~(Fig.~\ref{2D_sensor}). Hamamatsu S11639-01 CMOS chips with a quartz cover window and a protective film on the pixels were used~\cite{hamamatsu_cmos_linear}. The sensor mount was 3D printed using stainless steel to minimize thermal expansion effects and increase mechanical stability compared to the materials used for plastic 3D printing. The coefficient of thermal expansion of the stainless steel can be 4-10 times lower, depending on the chosen plastic~\cite{Callister2020MaterialsScience, Ivanova2022ThermalPlastics}. The mount was equipped with laser tracker fiducial (reference) marks. A fiducialisation is a process that links the reference geometry of the sensor to the fiducial marks~\cite{CERN2024}. Each sensor was placed on a three-ball kinematic mount~\cite{Touze2010CLICMRN}. 

A so-called symmetry-breaking effect refers to a non-straight alignment reference given by the laser beam caused by the diffraction effects on the aperture~\cite{polak2022symmetry}. The aperture size of 75~mm was used for the first and second sensors. The last sensor had an aperture size of~40~mm. In the future, the sensors are intended for use on a 140~m long testing bench. The aperture sizes were chosen accordingly, to minimize the symmetry-breaking effects over the 140~m while keeping the beam visible on all the sensors. Note that these effects are negligible on the setup presented in this study as the distance is significantly shorter. 

\begin{figure}[!ht]
\centering
\includegraphics[width=0.8\textwidth]{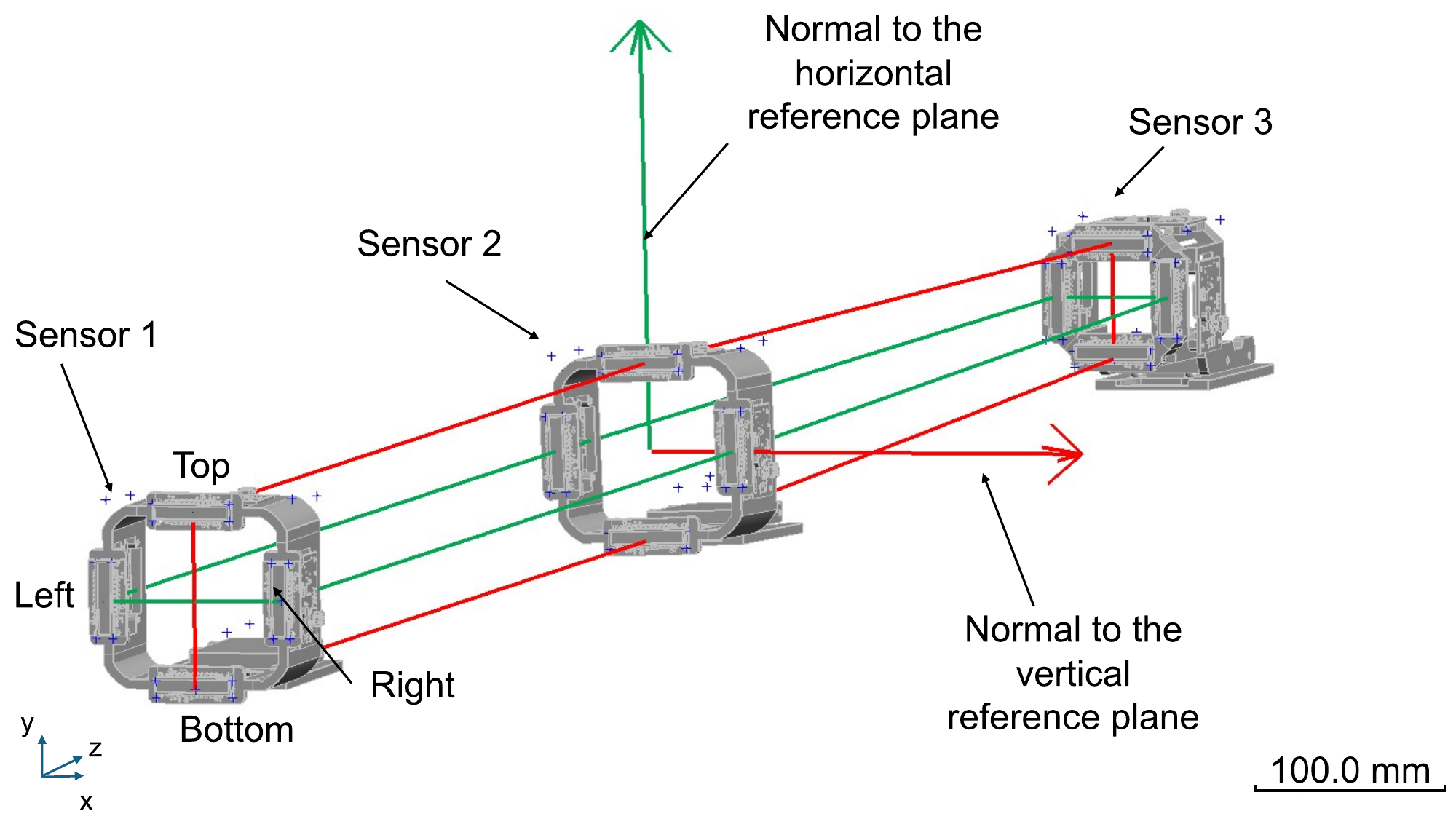}
\caption{Layout of the experiment. Two LBs generate the horizontal and vertical reference planes based on the readings from the CMOS chips. A photo of the experimental setup can be found in supplementary materials.}
\label{setup_model}
\end{figure}

To experimentally validate the performance of the developed sensors a dedicated measurement setup was established. Three sensors were arranged sequentially along the beam propagation path as illustrated in Fig.~\ref{setup_model}. The first one was positioned at a distance of 1.7~m from the LB generators. The spacing between the sensors was equal to 1~m resulting in a total length of 3.7~m from the generator to the last sensor. To mitigate issues arising from beam propagation through a non-homogeneous medium, a non-reflecting black pipe was placed in between the sensors. This minimized the refractive index fluctuation caused by the mixing of the air in the laboratory. In the future, we plan to propagate the LBs inside a vacuum pipe to minimize the disturbance caused by the non-homogeneous refractive index of the propagation medium.

\subsection{Fiducialisation}\label{fid_chap}
\hspace{5mm} Before installation in the experimental setup, each sensor underwent a metrological characterization in a dedicated laboratory. The positions of the active parts of the CMOS chips, fiducial marks, and three-ball mount interface within a sensor's reference frame were precisely determined in the local Cartesian coordinate system. This characterization was done utilizing coordinate-measuring machines~(CMM)~\cite{hocken2012coordinate}. Touch and optical probes were used, resulting in a measurement uncertainty of \SI{2.5}{\um}. These measurements provided accurate coordinates of the pixel positions relative to the fiducial targets and to the standardized three-ball mount interface, ensuring the common local coordinate frame within each sensor.

After placing the sensors in the experimental setup a total of 24 fiducial marks (8 on each sensor) were measured using a Leica AT401 laser tracker from three instrument stations. The measurements were linked using the Unified Spatial Metrology Network (USMN) module within the commercially available Spatial Analyzer metrological software. This allowed for the integration of all observations into a single measurement network. The network achieved an overall RMSE of \SI{0.004}{\milli\meter}, with a maximum point residual of \SI{0.015}{\milli\meter}.

To relate the local sensor coordinate systems to the global coordinate system obtained from the laser tracker measurement, best-fit transformations were performed between the fiducial marks measured by the CMM and laser tracker. This allowed us to determine the absolute position of all the active parts of the CMOS chips, fiducial marks, and three-ball mount interfaces within a common global coordinate system. The results are summarized in the Tab.~\ref{tab:bestfit_results}. The values of the RMSE for all the fits were under \SI{0.013}{\milli\meter} with a largest point residual value equal to \SI{0.017}{\milli\meter}. The global Cartesian coordinate system was chosen to be the same as the one used by the first laser tracker station and the results presented are within this coordinate system.  

\begin{table}[!ht]
\caption{\label{tab:bestfit_results} Best-fit transformation results between each sensor's local coordinate system and the global USMN frame.}
\begin{indented}
\item[]\begin{tabular}{@{}llll}
\br
\textbf{Metric} & \textbf{Sensor~1~[mm]} & \textbf{Sensor~2~[mm]} & \textbf{Sensor~3~[mm]} \\
\mr
RMSE & 0.009 & 0.012 & 0.006 \\
Maximum residual & 0.011 & 0.017 & 0.008 \\
\br
\end{tabular}
\end{indented}
\end{table}

\subsection{Characterization of the alignment system}
\hspace{5mm} Ideally, an LB should create a perfect reference plane, ensuring that all 6 points measured on all 3 sensors lie precisely on this plane. In reality, measurement uncertainty from the ideal plane arises due to various factors such as the dark noise of the CMOS chips, atmospheric disturbances, and interference on both: the CMOS cover glass, and the protective film of the pixels. Assessing the quality of this reference plane is essential because it directly impacts the precision of alignment measurements. 

Once the LB is incident on the sensors, the signal on each CMOS chip was analyzed using a weighted center of gravity algorithm (first moment weighted by intensity) with gamma correction~\cite{CentroidPaper}. This allowed for a calculation of a centroid coordinate with a sub-pixel resolution. The gamma correction was applied to the intensity vector $\mathit{I(x)}$ before calculating the centroid coordinate:

\begin{equation}
G_{corr}(x) = I_{max} \left(\frac{I(x)}{I_{max}}\right)^\gamma
\end{equation}

where $I_{max}$ is the maximum intensity value. The value of $\gamma = 8$ was used. The centroid coordinate~$\mathit{\overline{x}}$ of the intensity vector was then calculated using the center of gravity algorithm as:

\begin{equation} 
\label{centroid}
\overline{x} = \frac{\sum_{x=0}^{x_{max}}  G_{corr} (x) x}{\sum_{x=0}^{x_{max}} G_{corr} (x)}
\end{equation}

This approach gave the absolute position of 6 points in the global coordinate system for both, horizontal and vertical directions. A plane was fitted among these 6 points and the error of the fit was analyzed. Several experiments were conducted on the performance of the system:

\subsubsection{Temporal repeatability:}\label{temporal_chap}
The temporal repeatability of the quality of the alignment reference generated by the LBs was measured by performing twenty consecutive measurements for each plane spaced by one-second intervals, without altering any system parameters.

An offset for each point from the ideal fitted plane was calculated. The RMSE was determined from these offsets. This resulted in twenty values of the RMSE, one for each measurement. The mean value of the twenty RMSEs was calculated with a confidence interval of 2-sigma. Additionally, a maximum offset value of the point from the ideal reference plane was identified to illustrate the largest deviation. The same metrics for error evaluation were employed to analyze the results in the following sections~\ref{diff_planes_chap}~and~\ref{reproducibility}.

\subsubsection{Repeatability for different reference planes:}\label{diff_planes_chap}
A high-quality parallel optical window was introduced into the beam path and rotated between each of the twenty measurements, while the sensors were not moved. A single rotation induced controlled lateral shift of \SI{50}{\um} in the reference plane position. This allowed us to assess whether the quality of the alignment reference remained consistent for different reference plane positions.

Twenty measurements were taken for each position of the plane. The measurements were not taken simultaneously for both reference planes because the glass plate had to be rotated by 90~degrees around the optical axis between the measurements to shift the position of either the vertical or horizontal plane. It was ensured that the angle was not significantly changed with respect to the cover glass of the CMOS chips.

\subsubsection{Reproducibility:}\label{reproducibility}
An additional experiment was conducted to evaluate the robustness of the alignment system under different sensor orientations and reproducibility. The sensors were rotated by 90~degrees around the z-axis. This rotation effectively swapped the CMOS chips in detecting horizontal and vertical reference planes. The absolute sensor positions were remeasured using a Leica AT403 laser tracker with similar measurement errors of the USMN and of the best fit as presented in the section~\ref{fid_chap}. Again as in the case of the temporal repeatability experiment in the section~\ref{temporal_chap}, twenty consecutive measurements were spaced by one-second intervals without altering any parameters. This measurement was conducted several days after the aforementioned experiment. 

\subsubsection{Relative alignment:} Finally, the relative alignment capabilities of the system were tested. The reference plane was defined based on the known positions of the first and last sensors (Sensors~1~and~3) obtained by the laser tracker measurement. The middle sensor (Sensor~2) was mounted on a motorized linear stage and displaced repeatedly by precisely \SI{0.050}{\mm} increments twenty times in the horizontal direction. Each displacement of the middle sensor was measured relative to the reference planes defined by the points measured on the first and the last sensors. This allowed us to perform an analysis of the relative linear misalignment of the middle sensor with respect to the alignment reference given by the LBs. The mean detected displacement was calculated within a 2-sigma confidence interval, together with a standard deviation, and minimum and maximum detected misalignment. 

This measurement aimed to simulate the foreseen LBs-based alignment system, where reference planes will be defined by the first and last sensors. The sensors in between will be aligned relative to these reference planes. Note, that the orientation of the sensors was the same as in the experiments described in sections~\ref{temporal_chap}~and~\ref{diff_planes_chap}.

\subsection{Radiation hard sensor}
\hspace{5mm} Electronic components like CMOS chips are susceptible to radiation damage when used in accelerator environments. To address this, we propose using a fiber matrix system to transmit the signal away from the radiation zone. This would protect the electronic components as they can be shielded away from the radiation. However, a key consideration is how the fiber matrix affects image quality and signal intensity, which can impact the detection accuracy of the reference plane. 

\begin{figure}[!ht]
\centering
\includegraphics[width=0.8\textwidth]{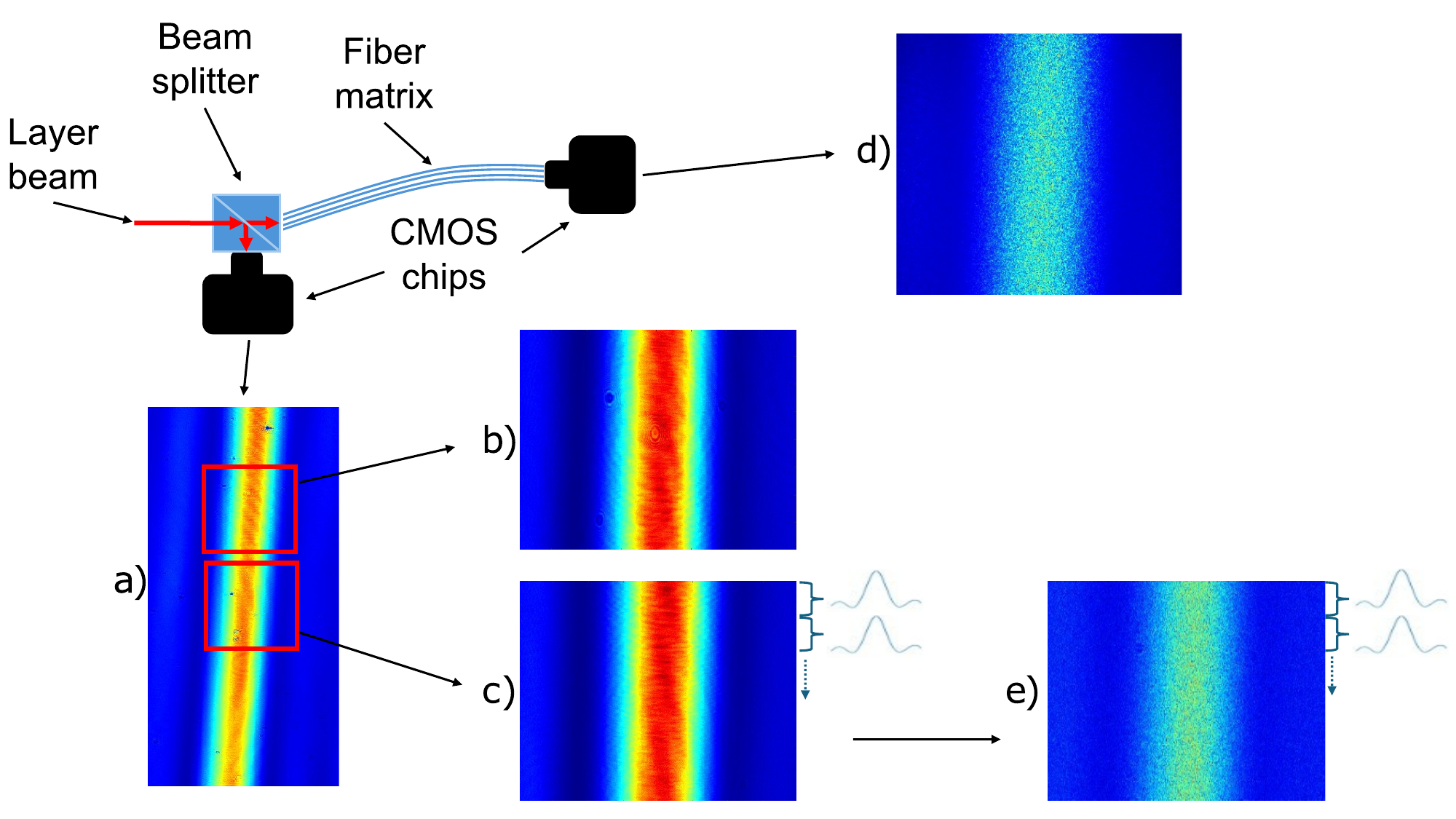}
\caption{Top view of the measurement setup for the analysis of the influence of the optical fiber matrix on the signal. a) Transversal intensity distribution seen on the CMOS without fiber matrix, b) extracted images for PSNR baseline comparison c) extracted image for LB signal analysis, d) Transversal intensity distribution after propagating through the fiber matrix, e) [Image c)] with induced noise for LB signal analysis.}
\label{fiber_measurement}
\end{figure}

A fiber matrix made by Schott, measuring 1050~mm in length was used in this experiment. The shape of the matrix is round with a diameter of 3.175~mm. It consists of a matrix of individual fibers each with a diameter of \SI{12}{\um}. A Layer beam was propagated over 8~m and then split by a beam splitter. A generator with a projection lens focal length of 100~mm was used for this experiment. On one side of the beam splitter, a camera Basler a2A5328-15umPRO was placed without any imaging optics. This camera features a CMOS chip with a resolution of 5328×4608 pixels and a pixel size of 2.74×2.74~\SI{}{\um}. On the other side of the beam splitter, a fiber matrix was put with the same camera type behind it. The end of the fiber matrix was positioned close to the CMOS chip, and the signal from the matrix was incident directly on it. The experimental layout is illustrated in Fig.~\ref{fiber_measurement}. 

Two equally sized images (400x800 pixels) were extracted from the transversal intensity distribution captured directly by the CMOS chip (Fig.~\ref{fiber_measurement}(b)~and~(c)). These images were compared using the Peak Signal-to-Noise Ratio (PSNR) method~\cite{Korhonen}. Since both images originated from the same intensity distribution, the PSNR value was relatively high, providing a baseline for quality assessment.

As a next step, an image from the camera placed behind the fiber matrix containing the LB signal (Fig.\ref{fiber_measurement}(d)), was compared to the image captured directly by the CMOS chip (Fig.\ref{fiber_measurement}(c)) using the PSNR. Since both images contain signals from the same LB, this comparison aimed to quantify the image quality and noise introduced by the transmission through the fiber matrix.

To simulate the insertion of the fiber matrix, the speckle noise, which would result in the same PSNR value as after the transmission through the fiber matrix, was induced to the image detected directly behind the beam splitter (Fig.\ref{fiber_measurement}(c)~and~(e)). The corresponding standard deviation of the Gaussian was calculated using the formula using the PSNR value:

\begin{equation}
\sigma = \sqrt{\frac{I_{\text{max}}^2}{10^{\frac{\text{PSNR}}{10}}}}
\end{equation}

Where $I_{\mathrm{max}}$ is the maximum pixel intensity of the normalized grayscale image. A zero-mean Gaussian noise $N(0, \sigma^2)$ was introduced into the image. A scaling function was used to enhance the noise presence in bright regions. The dynamic range of the image was scaled. This allowed us to simulate an image (Fig.~\ref{fiber_measurement}(e)) that would match the noise present in the image captured after propagating through the fiber matrix (Fig.\ref{fiber_measurement}(d)). 

As illustrated in Fig.~\ref{fiber_measurement}(c)~and~(e), from both, noise-induced and original image, twenty rows were summed up along the y-axis to simulate a larger pixel size of \SI{27.2}{\um}. This was done to increase the signal-to-noise ratio and to simulate a linear CMOS chip with non-square pixels. This process was repeated and resulted in twenty signals of the LB from each image that were processed using the equation in Eq.~\ref{centroid}. The differences in the coordinates between twenty signals from both images were compared using the RMSE with a 2-sigma confidence interval and maximum detected error. 

\section{Results}
\subsection{Layer beam-based alignment system}

\hspace{5mm} Several noise-inducing effects were discovered while conducting the experiments. If the LB direction of propagation was not perpendicular to the glass cover window of the CMOS chip, the signal was strongly distorted or asymmetrical most likely due to refraction and multiple internal reflections causing interference effects. 

\begin{figure}[!ht]
\centering
\includegraphics[width=\textwidth]{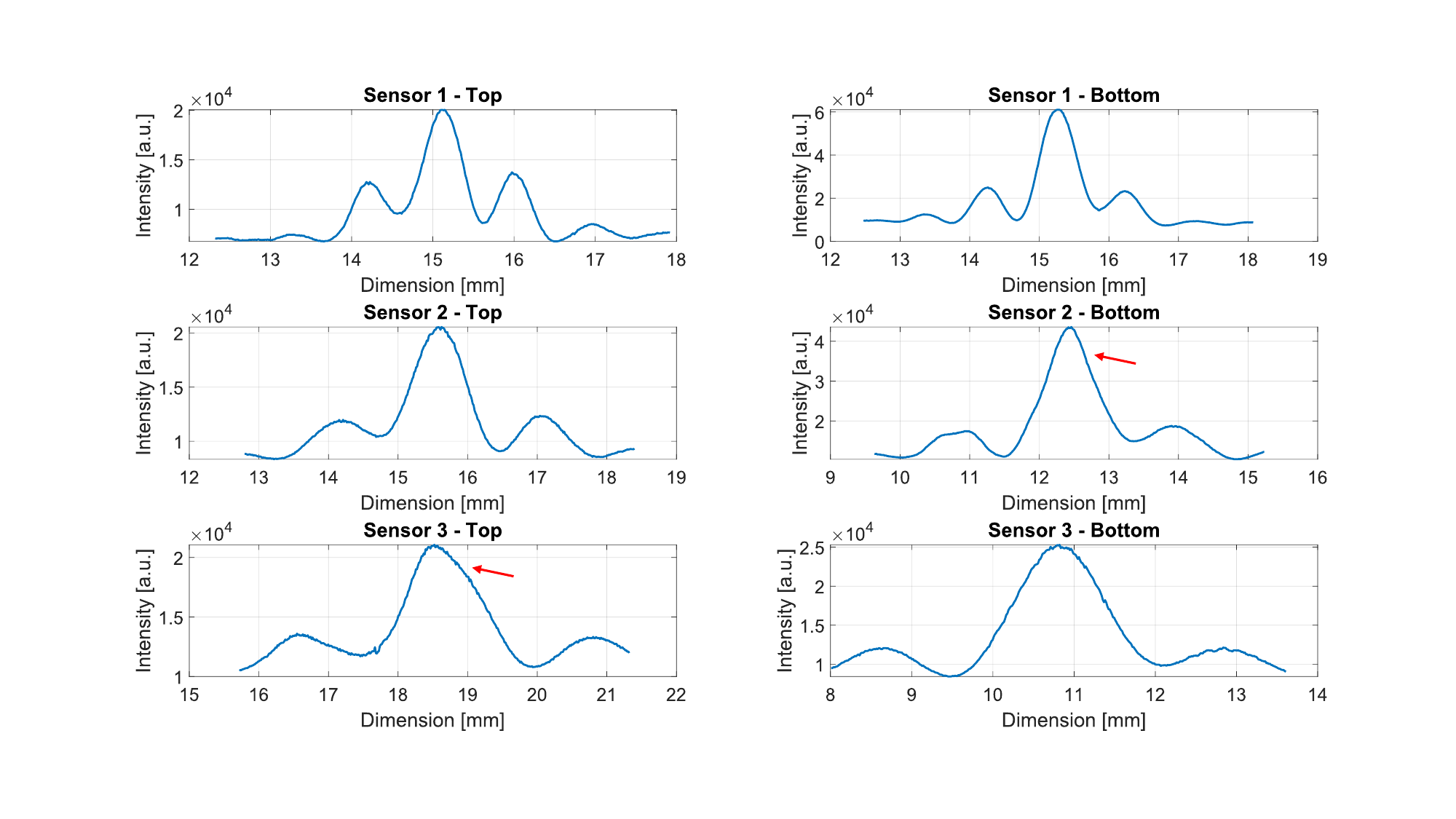}
\includegraphics[width=\textwidth]{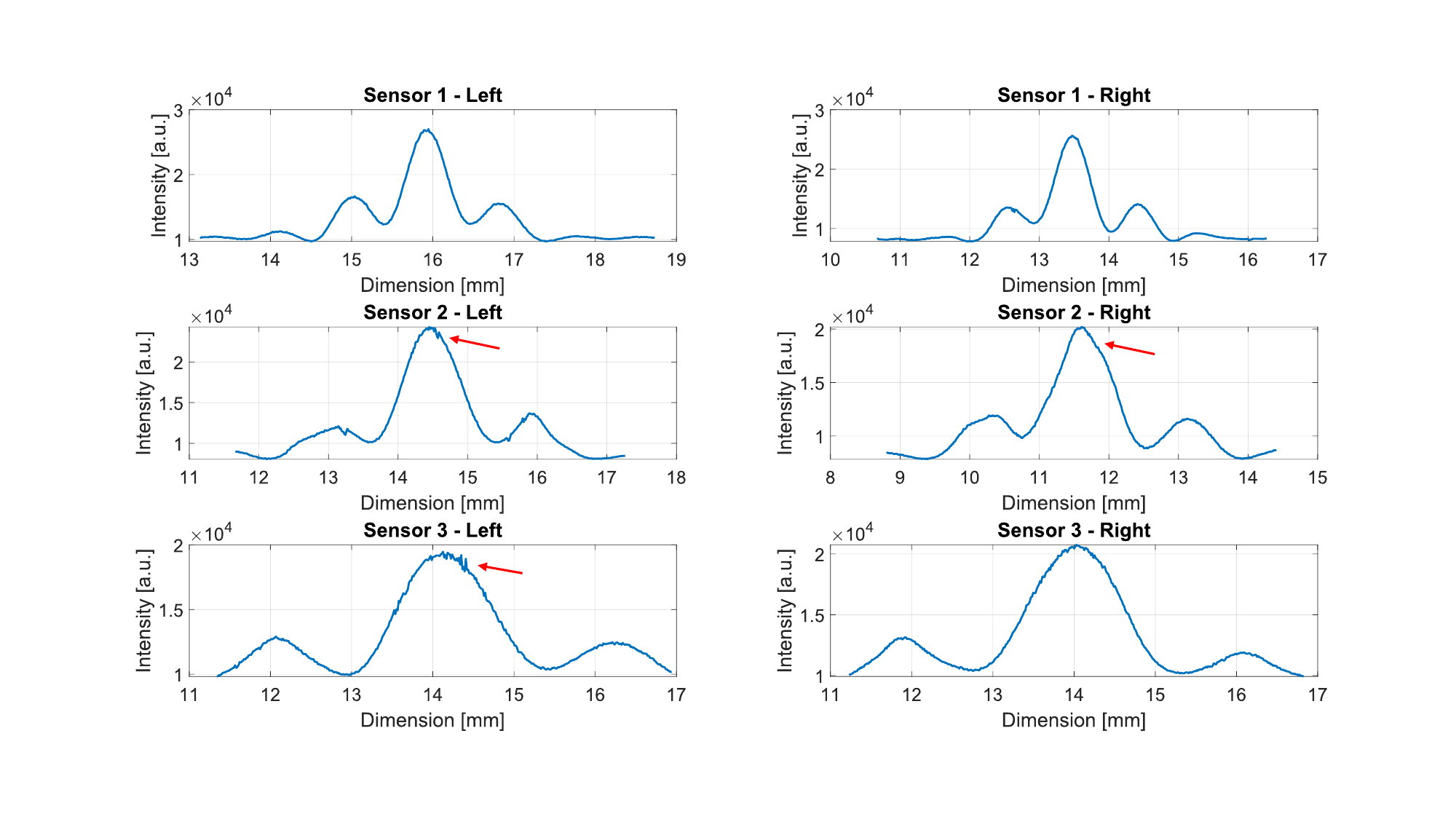}
\caption{Signals visible on each CMOS chip. Red arrows are pointing at the regions of the signal significantly disturbed by noise.}
\label{signal_example}
\end{figure}

A significant time instability of the signal at each CMOS chip was evident. One factor causing this was the background electronic noise (dark noise) of the CMOS chips visible during the measurements. An additional source was the atmospheric turbulence together with a temperature gradient causing local changes in the refractive index of the propagation medium. To characterize this behavior the temporal stability of the detected centroid points, obtained from the signal on each individual CMOS chip, was assessed by performing twenty measurements spaced one second apart. The results, summarized in Tab.~\ref{sensors_tab}, were analyzed using a standard deviation and minimum-maximum difference of the detected positions. The results indicate that the fluctuations are the lowest for Sensor~1, increase for Sensor~2, and are the highest for Sensor~3.

Measures, such as increased exposure time, background removal, and careful alignment of the reference plane with respect to the CMOS cover glass were employed. However, none of these effects could be completely eliminated during measurement, necessitating post-processing of the signal using low-pass filtering. The example of signals visible on each of the CMOS chips of every sensor can be seen in Fig.~\ref{signal_example} with red arrows indicating the asymmetric or noisy signals.

\begin{table}[!ht]
\centering
\caption{\label{sensors_tab} Results of temporal repeatability of the centroid point position obtained from the LB signal at each CMOS chip.}
\begin{indented}
\item[]\begin{tabular}{@{}lccc@{}}
\br
\textbf{} & \textbf{Sensor~1~[mm]} & \textbf{Sensor~2~[mm]} & \textbf{Sensor~3~[mm]} \\
\mr
\textbf{Top CMOS chip} & & & \\
\quad Standard deviation & 0.0005 & 0.0020 & 0.0049 \\
\quad Min-Max & 0.0020 & 0.0067 & 0.0149 \\
\mr
\textbf{Bottom CMOS chip} & & & \\
\quad Standard deviation & 0.0004 & 0.0015 & 0.0038\\
\quad Min-Max & 0.0018 & 0.0054 & 0.0135\\
\mr
\textbf{Left CMOS chip} & & & \\
\quad Standard deviation & 0.0004 & 0.0012 & 0.0042 \\
\quad Min-Max & 0.0019 & 0.0049 & 0.0143\\
\mr
\textbf{Right CMOS chip} & & & \\
\quad Standard deviation & 0.0005 & 0.0021 & 0.0033\\
\quad Min-Max & 0.0020 &  0.0070 & 0.0131\\
\br
\end{tabular}
\end{indented}
\end{table}

\subsubsection{Temporal repeatability} 
The results of the temporal stability experiment, presented in Tab.~\ref{temporal}, indicate a higher variability in the horizontal plane compared to the vertical plane. Specifically, the horizontal plane exhibited a mean RMSE of $0.025~\pm~0.002$~mm, while the vertical plane had a lower value of $0.017~\pm~0.002$~mm. The maximum offset value was also larger for the horizontal plane ($0.050$~mm) than for the vertical plane ($0.045$~mm). 

\begin{table}[!ht]
\centering
\caption{\label{temporal} Results of temporal repeatability.}
\begin{indented}
\item[]\begin{tabular}{@{}lll}
\br
\textbf{Metric} & \textbf{Vertical plane [mm]} & \textbf{Horizontal plane [mm]}\\
\mr
Mean RMSE & 0.017 ± 0.002 & 0.025 ± 0.002\\
Mean RMSE interval & [0.008, 0.025] & [0.019, 0.029] \\
Maximum point offset & 0.045 & 0.050 \\
\br
\end{tabular}
\end{indented}
\end{table}

\subsubsection{Repeatability for shifted reference planes}
The results of the repeatability analysis for different reference planes are presented in Tab.~\ref{shift_table}. A higher variability in the horizontal plane compared to the vertical plane can be observed again. The values of the RMSE have increased compared to the temporal stability by $0.004$~mm and $0.002$~mm for the vertical and horizontal planes respectively.

\begin{table}[!ht]
\centering
\caption{\label{shift_table} Results of repeatability analysis for shifted reference planes.}
\begin{indented}
\item[]\begin{tabular}{@{}lll}
\br
\textbf{Metric} & \textbf{Vertical plane [mm]} & \textbf{Horizontal plane [mm]}\\
\mr
Mean RMSE & 0.021 ± 0.002 & 0.027 ± 0.002\\
Mean RMSE interval & [0.016, 0.026] & [0.019, 0.030] \\
Maximum point offset & 0.044 & 0.052 \\
\br
\end{tabular}
\end{indented}
\end{table}

\subsubsection{Reproducibility}

The results of the temporal stability experiment after rotating the sensors by 90~degrees are presented in Table~\ref{temporal90}. Compared to the aforementioned temporal stability measurement, a shift in variability between the horizontal and vertical planes can be observed. The mean RMS for the vertical plane increased to $0.024~\pm~0.002$~mm, while the value for the horizontal plane decreased to $0.020~\pm~0.002$~mm.

\begin{table}[!ht]
\centering
\caption{\label{temporal90} Results of temporal repeatability for 90~degrees rotated sensors.}
\begin{indented}
\item[]\begin{tabular}{@{}lll}
\br
\textbf{Metric} & \textbf{Vertical plane [mm]} & \textbf{Horizontal plane [mm]}\\
\mr
Mean RMSE & 0.024 ± 0.002 & 0.020 ± 0.002\\
Mean RMSE interval & [0.020, 0.026] & [0.016, 0.026] \\
Maximum point offset & 0.048 & 0.043 \\
\br
\end{tabular}
\end{indented}
\end{table}

\subsubsection{Relative alignment} 
The results of the relative alignment capability tests, presented in Table~\ref{relative_alignment}, indicate that the mean detected displacement closely matched the known increment of \SI{0.050}{\mm} induced by the motorized linear stage. A mean value of $0.051~\pm~0.002$ resulting from the twenty measurements was detected. The relatively high standard deviation of \SI{0.004}{\mm} with the maximum and minimum detected misalignment of \SI{0.056}{\mm} and \SI{0.042}{\mm} respectively suggest notable variability between the measurements. 

\begin{table}[!ht]
\caption{\label{relative_alignment} Results of the relative misalignment of the Sensor~2 for the known displacement of \SI{50}{\um}.}
\begin{indented}
\item[]\begin{tabular}{@{}ll}
\br
\textbf{Metric} & \textbf{Value [\SI{}{\mm}]} \\
\mr
Mean detected displacement & $0.051~\pm~0.002$\\
Standard deviation & 0.004\\
Maximum detected displacement & 0.056\\
Minimum detected displacement & 0.042\\
\br
\end{tabular}
\end{indented}
\end{table}

\subsection{Radiation-hard sensor concept}

\begin{figure}[!ht]
    \centering
    \begin{subfigure}{\columnwidth}
        \centering
        \includegraphics[trim=1cm 5cm 1cm 5cm,clip,width=\textwidth]{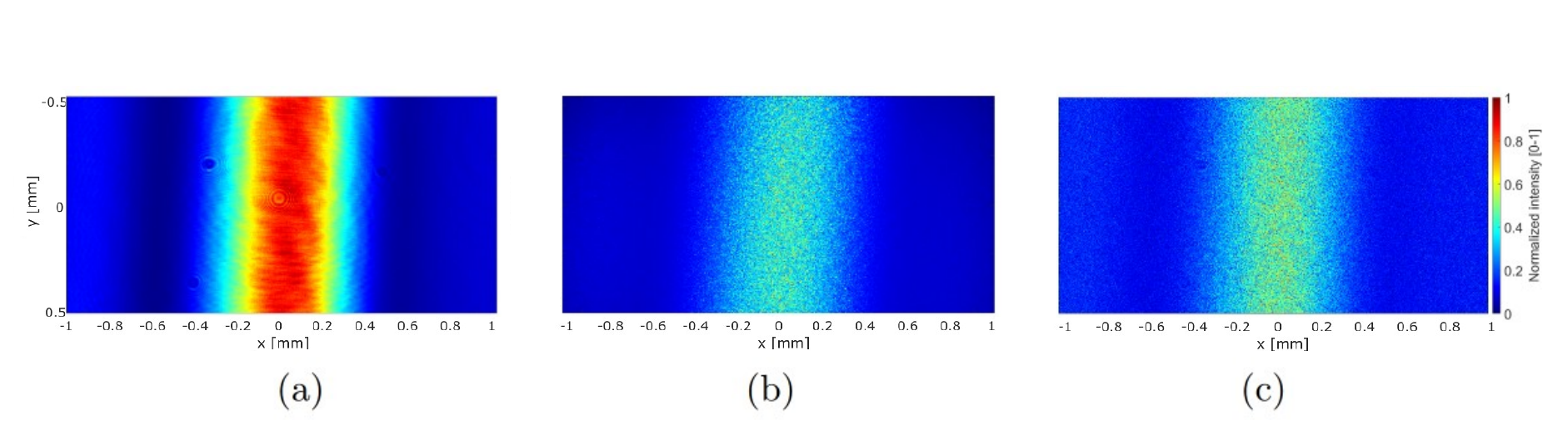}
    \end{subfigure}
\caption{Measured transversal intensity distributions of a single line (a)~without fiber matrix, (b)~after the fiber matrix, (c)~with induced noise.}    
\label{one_line}
\end{figure}

\begin{figure}[!ht]
    \centering
    \begin{subfigure}{\columnwidth}
        \centering
        \includegraphics[trim=1cm 5cm 1cm 5cm,clip,width=\textwidth]{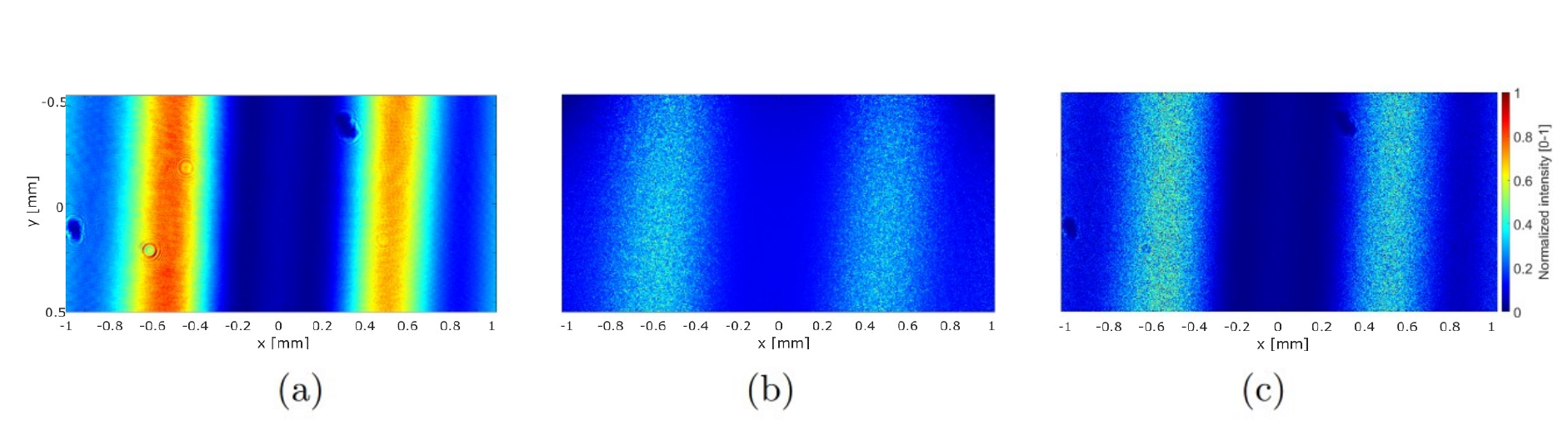}
    \end{subfigure}
\caption{Measured transversal intensity distributions of two lines (a)~without fiber matrix, (b)~after the fiber matrix, (c)~with induced noise.}   
\label{multiple_lines}
\end{figure}

\hspace{5mm} Tests were conducted on two beam patterns: one with a single line (Fig.~\ref{one_line}) and the other with two lines (Fig.~\ref{multiple_lines}), with the results summarized in the Tab.~\ref{fiber_table}. The comparison showed that the PSNR was notably affected by the fiber matrix, indicating a significant impact on the signal-to-noise ratio in the image.

For the single-line pattern, the PSNR decreased from the baseline value of 33.46~dB to 19.75~dB after fiber transmission. Analysis of the extracted LB signals showed corresponding values of RMSE equal to $1.24 \pm 0.42$~\SI{}{\um} and of a maximum error of \SI{2.72}{\um}. For the multi-line pattern, the PSNR decreased from 34.87 dB to 14.55 dB, with a corresponding RMSE of $0.71 \pm 0.25$\SI{}{\um} and a maximum error of \SI{2.19}{\um}.

\begin{table}[!ht]

\centering
\caption{\label{fiber_table} The results of how the fiber matrix influences the detection accuracy of the reference plane.}
\begin{indented}
\item[]\begin{tabular}{@{}*{3}{l}}
\br
\textbf{Metric} & \textbf{One line} & \textbf{Two lines} \\ \mr
Baseline PSNR [dB] & 33.46 & 34.87 \\ \mr
PSNR [dB] & 19.75 & 14.55 \\ \mr
RMSE [\SI{}{\um}] & $1.24 \pm 0.42$ & $0.71 \pm 0.25$ \\ \mr
Maximum error [\SI{}{\um}] & 2.72 & 2.19 \\
\br
\end{tabular}
\end{indented}
\end{table}

\section{Discussion}

\subsection{Layer beam-based alignment system}
\hspace{5mm} Preliminary results suggest that the LBs-based alignment system could be a viable candidate for future alignment applications. However, the experimental results revealed deviations from ideal alignment reference planes. This was primarily caused by the analysis of the non-ideal signals which were distorted by the combination of dark noise together with reflection and refraction effects on the cover glass and pixel protective film of the CMOS chips. The interference, due to multiple reflections on the cover glass and on the protective pixel layer, seems to be the main limiting factor. Therefore, all the tests were done only for relatively limited cases, where the LBs propagated as perpendicularly as possible to the CMOS chips to mitigate these effects. 

Additional distortion was caused by the non-homogeneous refractive index of the propagation medium. The fluctuations of the centroid coordinate in time, detected at each individual sensor, increase with the distance from the LB generator. This is most likely because the beam travels in the medium with a turbulent and non-homogeneous refractive index for a longer time. Additionally, while fiducialisation plays a critical role in aligning the sensor's local coordinate system to the global reference, it introduces errors. These errors can lead to systematic offsets in the reference plane, affecting the overall alignment accuracy. However, the results suggest that the noise induced by sensors and environmental fluctuations are the dominant sources of uncertainty in the alignment system compared to fiducialisation. This problem should be largely mitigated by propagating inside a vacuum pipe. Despite these challenges, the system showed promising repeatability and relative alignment capabilities.

While assessing the quality of the reference plane, the temporal stability experiment highlighted differences between the vertical and horizontal reference planes. The vertical plane exhibited better stability compared to the horizontal plane. This discrepancy, together with reproducibility, was further investigated by rotating the sensors by 90~degrees, effectively swapping the CMOS chips responsible for detecting each plane. The rotation reversed the observed trend, suggesting that signal distortion by the CMOS chips is a major source of measurement error further supporting the need for perpendicularity between the beam path and the cover glass of the CMOS chip. An additional factor in this could be the fiducialisation procedure. 

Controlled shifts of the reference planes using the parallel plate window confirmed the system's repeatability but again revealed sensitivity to slight changes in beam angle relative to the sensors. These findings underline the importance of minimizing interference and refraction effects. In the future, CMOS chips will be upgraded. Different types of linear chips need to be tested, specifically chips: without a cover window, with anti-reflective coating on the cover window, and also without the protective film on the pixels. Additionally, integrated dark noise correction should be employed. 

Relative alignment tests demonstrated that the system could detect known displacements induced by a linear motor, achieving a mean detected displacement close to \SI{50}{\um}. Again, this error can be primarily attributed to the instability of the reference plane in time caused by the refractive index gradient and fluctuation in the propagation medium due to environmental perturbations such as temperature. 

\subsection{Radiation-hard sensor}
\hspace{5mm} The experimental results demonstrate the potential feasibility of using fiber matrix systems for radiation-hard sensors in accelerator environments. Integrating a fiber matrix system into the sensor design could effectively protect electronics from radiation-induced damage while maintaining alignment accuracy comparable to direct CMOS detection.

The fiber matrix successfully transmitted the LB signal while having a relatively low impact on the accuracy of the reference detection. Comparative analysis of the transversal intensity distributions before and after propagation through the fiber matrix revealed that the PSNR gets strongly affected. However, it does not have a significant impact on the analysis of the LB signal. For both single-line and multi-line beam patterns, the RMSE remained below \SI{1.30}{\um}, with a maximum error of \SI{2.72}{\um}.

The comparison between the single-line and two-line beam patterns revealed that the two-line configuration provided a lower RMSE. The centroid calculation most likely becomes less sensitive to localized noise, leading to improved stability and precision in reference plane detection. Thus, employing a multi-line beam structure could enhance noise resilience.

\section{Conclusion}

\hspace{5mm} The key advantage of the LBs-based alignment system is its ability to measure multiple points simultaneously along its propagation path without obstructing the beam. Compared to the state-of-the-art straight-line alignment systems, an additional degree of freedom (the roll) could be measured. The experiments were performed over a relatively short distance of 2~m between the first and the last sensor. The quality of the reference plane generated by the LBs remained stable, with repeatability and reproducibility RMSE values below \SI{30}{\micro\meter}. The relative alignment measurements achieved repeatability with a standard deviation of \SI{4}{\micro\meter}.  

Lessons learned from these experiments highlight the need for upgrading CMOS chips to address issues related to light refraction and interference on the cover glass window and on the protective film of the pixels. The dark noise also needs to be reduced.

The system could be a viable alternative to the straight-line reference optical systems that rely on mechanical shutters. In the future, the system based on LBs will be compared to the one based on SLB on a 140~m long test bench. This setup includes a vacuum pipe inside which the beams will propagate. The results will be also compared with the state-of-the-art alignment methods WPS and HLS. 

The radiation-hard sensor concept, which utilizes the fiber matrix, introduced an additional RMSE below \SI{1.3}{\micro\meter} confirming its potential for use in high-radiation environments. In the future, a sensor with integrated fiber matrices needs to be developed and employed in a real alignment scenario for further analysis. Additionally, effects such as fiber degradation over time caused by radiation need to be investigated~\cite{Girard2019}.

\section*{Data availability statement}
\hspace{5mm} The data that support the findings of this study are available upon reasonable request from the authors.

\ack

\hspace{5mm} The authors acknowledge the financial support provided by the Knowledge Transfer group at CERN through the KT Fund. This work was partly supported by the Student Grant Scheme at the Technical University of Liberec through project SGS-2025-3583. 

\section*{Conflict of interest}
\hspace{5mm} The authors declare that they have no known competing
financial interests or personal relationships that could have appeared to influence the work reported in this study.

\section*{References}
\bibliographystyle{unsrt}
\bibliography{refined_full_references}
\end{document}